*Lightly edited transcript of lecture given at the "Standard Model at 50" Symposium held at Case Western University, June 1–4, 2018. The official Proceedings should eventually be published by Cambridge University Press; however, since this has not yet occurred after several years of delays, the lecture is now being made available on arXiv; it includes a brief update on the Hubble tension, at the end.*

**Expansion of the Universe** (Alexei V. Filippenko, University of California, Berkeley)

**[Corbin Covault]** The next speaker will be Alex Filippenko. He is a Professor of Astronomy, and the Richard and Rhoda Goldman Distinguished Professor in the Physical Sciences, at the University of California, Berkeley. He received his Bachelor of Arts in physics from UC Santa Barbara in 1979 and his PhD in astronomy from Caltech in 1984. A major area of Alex's work is the understanding of supernovae and their application to a range of problems including cosmology. His team developed a robotic telescope at Lick Observatory that has discovered more than 1,000 nearby supernovae. Alex is the one person who was a member of both teams that revealed the accelerating expansion of the universe, a discovery that was honored with the 2011 Nobel Prize in Physics to the team leaders, as well as the 2007 Gruber Cosmology Prize and the 2015 Breakthrough Prize in Fundamental Physics to all team members. Alex has also made several outstanding contributions promoting physics and astronomy education and outreach, including coauthoring an introductory astronomy college textbook and appearing in numerous television documentaries. Today Alex is going to tell us about the expansion of the physical universe. Let's welcome Alex Filippenko.

**[Alex Filippenko]** Well, thank you, Corbin, for that very nice introduction. It's a pleasure to speak here at this august meeting of people, but mostly particle physicists, and so I'll give you more of an elementary talk on astrophysics and cosmology, detailing how the discovery was made of the accelerating expansion of the universe.

We've known for nearly a century that the universe is expanding based on Edwin Hubble's analysis of galaxy redshifts measured by Vesto Slipher, a person who gets far too little credit, and Hubble's own measurements of the distances of galaxies. And as I'll show you later, now with telescopes like the *Hubble Space Telescope (HST)*, we have a very good measurement of the current expansion rate of the universe. Figure 1 shows an idealized Hubble diagram with no scatter, no noise: log of distance *d* versus log of redshift, or recession speed *v*. The value of the Hubble constant $H_0$ is just the slope of the linear relationship, $v = H_0 d$, known as the Hubble law (or, giving proper credit to Georges Lemaître, the Hubble-Lemaître law).

Now if we plot the scale factor $a(t)$ of the universe as a function of time *t* in Figure 2 – first, I show the case of an empty universe ($\Omega_M = \rho_{ave}/\rho_{crit} = 0$) with no gravity, a straight line sloping up to the right – the Hubble constant is just a-dot or a-prime over a, evaluated right now: $(a'/a)_0$. It's just some number that corresponds to the slope of that line divided by the value of the scale factor at this time. But you could instead have a dense universe ($\Omega_M > 1$) that reverses its motion, like the trajectory of this apple that I toss upward from my hand – and right now, it would have the same Hubble constant, the same $a'/a$ value. Or you could have a medium-density flat universe, one whose average density corresponds to the critical density ($\Omega_M = 1$); in this case, of course, the recession speeds of galaxies asymptotically approach zero. Again, it would have the same value of $H_0$.

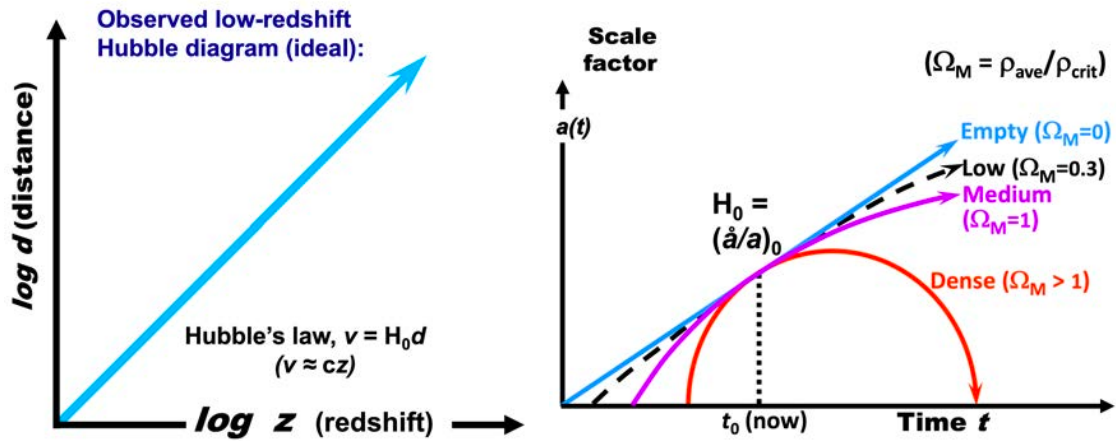

Figure 1 (left): Idealized low-redshift Hubble diagram. Figure 2 (right): Model universes having the same Hubble constant $H_0$.

In the early 1990s, observational astronomers measuring the distribution of galaxies and clusters of galaxies told us that we live in a relatively low-density universe, $\Omega_M \approx 0.3$, so the recession speeds of galaxies would asymptotically approach some constant value, but not zero. The question is, in which of these universes do we *really* live? Is there some other way to know?

Well, there *is* a way to know. We can look at the past history and thus predict the future. Let's zoom in on that part of Figure 2 representing the past. In Figure 3, I again have a*(t)* vs. *t*, and I've got the four cases: an empty universe, a low-density universe, a critical density universe, and a closed ($\Omega_M > 1$) universe. Here we are now, redshift zero. Redshift *z* is defined such that 1 + *z* is the scale factor of the universe now divided by the scale factor at the time that the photon we are now seeing was actually emitted. At redshift 1, for example, by advanced mathematics, 1 + 1 = 2; and 1 over 1/2 is 2, so the scale factor of the universe was half of what it is now when the photon from a galaxy at redshift 1 that we are now observing was actually emitted.

You can see here that the lookback time at fixed redshift depends on what the history of the universe has been. For a given redshift the lookback time, and hence the distance, is shortest in a dense universe, bigger in a medium-density universe, bigger still in a low-density universe, and the biggest possible in an empty universe

(or so we thought). So, by simply measuring the distances – and hence the lookback times – as a function of redshift, we should be able to tell the past history of the universe.

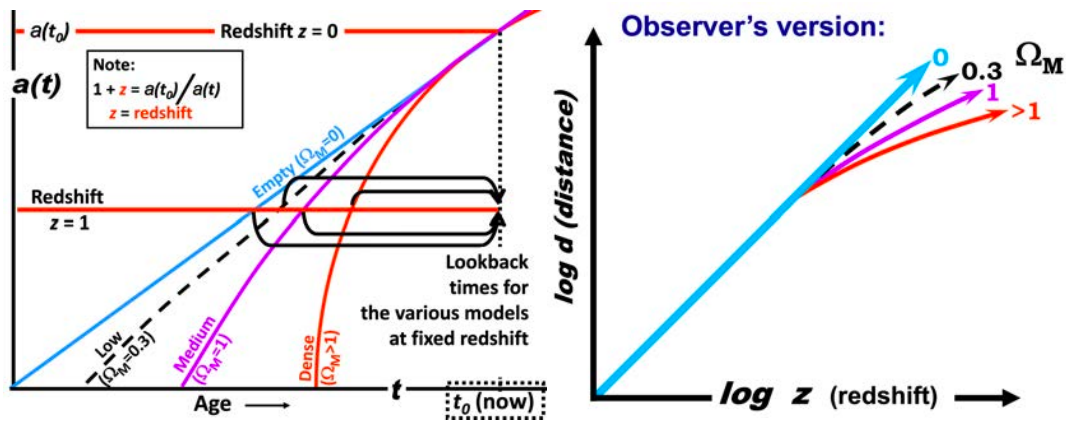

Figure 3 (left): Lookback times for various past histories; specific example given at redshift 1. Figure 4 (right): Observer's version of Figure 3.

Now, the *observer's* version of this diagram looks something like Figure 4, which shows log distance $d$ vs. log redshift $z$. The distance has to be determined by a method *other* than the redshift – it's not fair to use Hubble's law, because we're observing far enough back in time that we're looking precisely for *deviations* from that law. As you can see, the densest universe indeed has the smallest lookback time, the smallest distance for a given redshift. So, if you populate this diagram with observations, that will tell you the future of the universe, in principle.

Well, how do we determine the distances of these various galaxies? And how do we get the redshifts? The redshifts are easy: you just obtain the spectrum of a galaxy and you see spectral features, which you recognize in very nearby galaxies whose recession speed is essentially zero. Think about the spectrum of a nearby galaxy, as well as the same spectrum arbitrarily shifted by 1/10 the speed of light. It's the same spectral lines, so you just measure the wavelength $\lambda$ of a given line and compute $z = \Delta\lambda/\lambda_0$, where $\lambda_0$ is the rest wavelength of this particular line. Alright, so that's easy to do.

The distances are a lot harder to determine. I'm going to use distances that are called luminosity distances; there are many types of distances in cosmology, and you can get into trouble if you mix them. Luminosity distance is just the distance that exactly satisfies the inverse-square law of light. So, you measure a flux (an apparent brightness) of an object whose luminosity (or intrinsic brightness) you somehow already know, and that gives you the luminosity distance. The hard part is knowing the object's intrinsic luminosity. How do you do that? You need some sort of a standard candle. They don't exist, even though astronomers call lots of things "standard candles." They're actually *standardizable*, not standard.

In a nearby galaxy, it's easy to identify individual stars, at least the more luminous normal stars – especially in an *HST* image. And for nearly a century what's generally been used are the Cepheid variables. They're recognized by their light curves, apparent brightness vs. time. They rise (brighten) pretty steeply, then decline (fade) more slowly with time, a compete light-curve cycle being one period. And they have what's called a period-luminosity relationship – the average luminosity increases with increasing period. If you know the period-luminosity relationship, then measuring the period of a Cepheid tells you its average luminosity, and thus you've standardized your candle. I call them standardizable candles, not standard candles.

Anyway, this was first done for the so-called "spiral nebulae" by Edwin Hubble with the 100-inch Hooker telescope at Mt. Wilson Observatory, above the San Gabriel Valley. In October 1923, he famously discovered a faint Cepheid variable in the Andromeda nebula, showing that the nebula was actually an external galaxy, an "island universe," in a sense. That's not very hard to do in nearby galaxies, at least not these days. But how about faint, distant galaxies such as in the Hubble Ultra Deep Field, little smudges which are billions of light years away? Cepheids aren't luminous enough to be seen in such distant galaxies.

So, we use another technique – supernovae, which are exploding stars. They can become millions or even billions of times the power of our Sun, and there are several different types. The one that's most useful for cosmology are the Type Ia supernovae (SNe Ia). Their optical spectra show all kinds of elements, but notably no hydrogen nor any helium, as opposed to the Type II supernovae that do exhibit hydrogen. SNe Ia are thought to be nearly standard or standardizable, both observationally and theoretically. Theoretically they come from a white dwarf at or near the Chandrasekhar limit of around 1.4 solar masses. It can approach that mass by stealing material from a more-or-less normal companion star.

Or, a double white dwarf merger can occur. In this case, the less massive white dwarf gets tidally disrupted by the more massive one and forms an accretion disk. So, the thing should still explode near the Chandrasekhar limit, even though in principle the sum of the white dwarf masses could exceed 1.4 solar masses – but you get this dynamical accretion such that the SN Ia explosion still occurs at or near the Chandrasekhar limit.

Well, they're pretty standard, by they're not exactly standard. This was noticed decades ago, and Mark Phillips found an interesting relationship among SNe Ia studied in nearby galaxies: the intrinsically more luminous ones (at peak brightness) have slower light curves than those with normal (i.e., typical) peak luminosity (Phillips 1993). He measured the decline rate, but now we know that even the rise rate of luminous SNe Ia is slower than that of the normal ones; both of these trends are shown in Figure 5. And then there are subluminous SNe Ia, which rise and decline more quickly than the normal ones.

This technique was then perfected, primarily by Adam Riess in his doctoral thesis with William Press and Bob Kirshner at Harvard University (Riess et al. 1995, 1996) as well as by Mario Hamuy and collaborators in Chile (Hamuy et al. 1996a,b). By determining the relationship between peak luminosity and decline rate with a set of only about ten nearby SNe Ia in the mid-1990s and then applying it to other SNe Ia in galaxies of unknown distance, you could then figure out the distances of those galaxies. You use this technique to standardize the candle. The devil is in the details, though; you need to account for interstellar dust which can dim the SNe Ia for a reason other than distance. For example, if there's fog in the way, a car headlight will look dimmer than it would with no fog. So, you measure the colors of an SN Ia, and that gives you its reddening, from which you can determine the amount of extinction of light. It's like the setting Sun: it looks both dimmer and redder (yellow if there's not much dust, orange if there's more dust, or even red if there's a lot of dust or smoke).

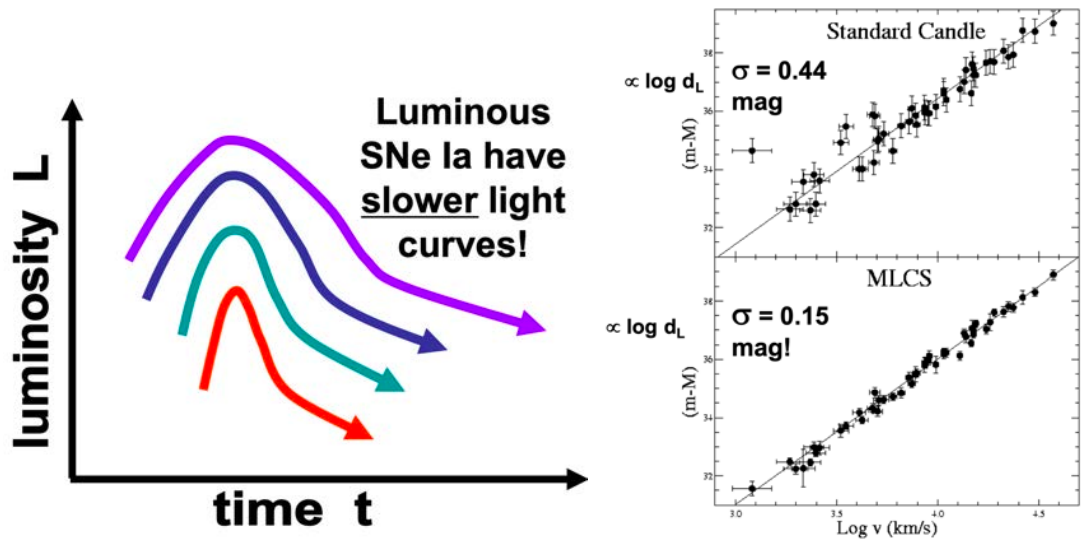

Figure 5 (left): Absolute light curves of Type Ia supernovae in galaxies of known distance. Figure 6 (right): Hubble diagram using SNe Ia as standard candles (top) and as standardizable candles (bottom); the scatter is much reduced in the latter.

When all is said and done, this multicolor light-curve shape method (Riess et al. 1996) works remarkably well. Figure 6 (left) shows the Hubble diagram assuming all SNe Ia are standard candles with some dispersion, say 50% or so (the "magnitude" unit is proportional to the logarithm of the brightness). If one then applies the multicolor light-curve shape technique to those same objects, as determined from a set of about a dozen nearby and well-observed SNe Ia, Figure 6 (right) shows that the dispersion decreases by a factor of 3, thereby making SNe Ia enormously useful yardsticks for cosmological studies. For example, the SN Ia with the smallest recession speed in Figure 6 (top) looks much too distant, and that's because it was behind a screen of dust. But after the correction is applied, boom, it just falls right where it should in Figure 6 (bottom) – it's fantastic!

Next, the trick was to go and find SNe Ia in lots of distant galaxies. Having calibrated their luminosities based on nearby galaxies, we could then get distances and hence the lookback times of the distant galaxies. So, in the early 1990s two teams were formed. Saul Perlmutter of the Lawrence Berkeley National Lab and UC Berkeley led one of them – the Supernova Cosmology Project – and Brian Schmidt and Nick Suntzeff formed the High-$z$ Supernova Search Team (Schmidt later became the official team leader). They weren't always at each other's throats; it was generally a healthy competition, and we occasionally talked to each other. For example, decades ago at the Aspen Center for Physics we discussed various ways of calibrating the SNe. Having two teams accelerated research progress, if you'll pardon the pun; both wanted to be first. It improved the quality of the science because both wanted to be best; if one team was taking a thorny technical issue into account and the other team was not, then that second team would look bad, in comparison. And finally, and most importantly, had there not been two teams that got the same results at nearly the same time, almost nobody would have taken our results very seriously.

Both teams used telescopes primarily in the southern hemisphere (specifically, the 4 m Blanco telescope at the Cerro Tololo Inter-American Observatory) to take wide-angle pictures of the sky that have thousands of galaxies in them. We repeated those regions three weeks later, digitally subtracted the old images from the new images, and found things that looked like they might be new. In Figure 7, cleverly placed near the center of the circle, is something that looks like it might be real among all the noise.

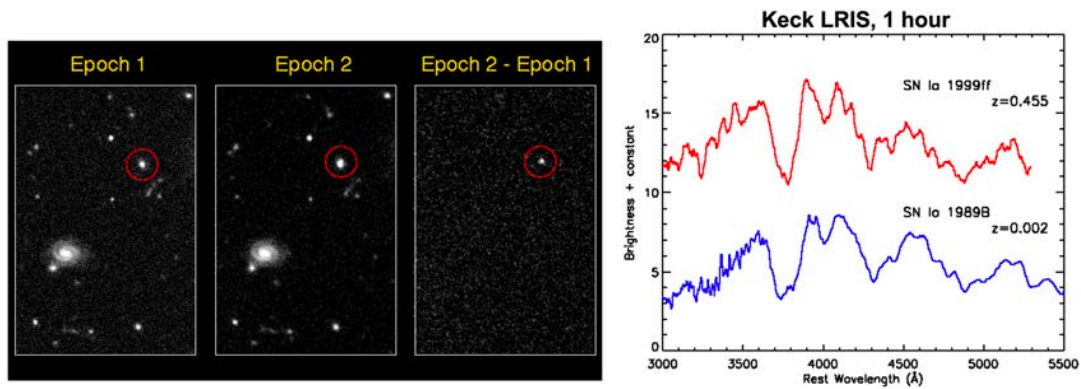

Figure 7 (left): Images obtained by the High-z Supernova Search Team, revealing a SN Ia candidate (circled) in the difference image. Figure 8 (right): Keck spectrum of SN 1999ff at $z$ = 0.455 compared with the spectrum of low-redshift SN 1989B.

My job on both teams was to then use the Keck 10 m telescopes to obtain spectra of these candidates to determine whether they really were SNe Ia, and to get the redshifts of the galaxies – remember, you need not only the luminosity distance but the redshift. I'm a trained spectroscopist; I've been working on supernovae since the mid-1980s, and I have privileged access to these wonderful, powerful telescopes.

Figure 8 shows a spectrum of one of the SN Ia candidates, SN 1999ff, at redshift 0.455. You can see that, to within the noise, it looks almost exactly like a redshift-zero SN Ia. Data like this made us very happy. Adam Riess, who by 1996 was a postdoctoral scholar (a Miller Fellow for Basic Research in Science) at UC Berkeley working with me, did the photometric analysis of the SNe Ia in the High-*z* Supernova Search Team's sample. (By this time, I was working primarily with Schmidt's team, although earlier I had contributed significantly to Perlmutter's team.)

Alright, so what did we find? Figure 9 shows three spectroscopically confirmed SNe Ia. They're faint – they look really faint. And you might say, "Well, what's the big deal? They're in these pathetic-looking faint galaxies. In one particular case you can't even see the galaxy. Obviously, these are very distant." And that's true, but the point is they're more distant than they should have been in any reasonably-behaved universe – or at least, any reasonably-behaved Newtonian universe. Figure 10 displays the possibilities that I outlined in Figure 4, and the data actually fell *above* all of the possibilities. In other words, at a given redshift the supernovae looked fainter and thus were more distant than they should have been in any standard cosmological model. Or course, we had to account for the presence of dust and the possible evolution of SNe Ia – maybe 5 billion years ago they were intrinsically less luminous than they are now. We tried to take that into account, as best we could.

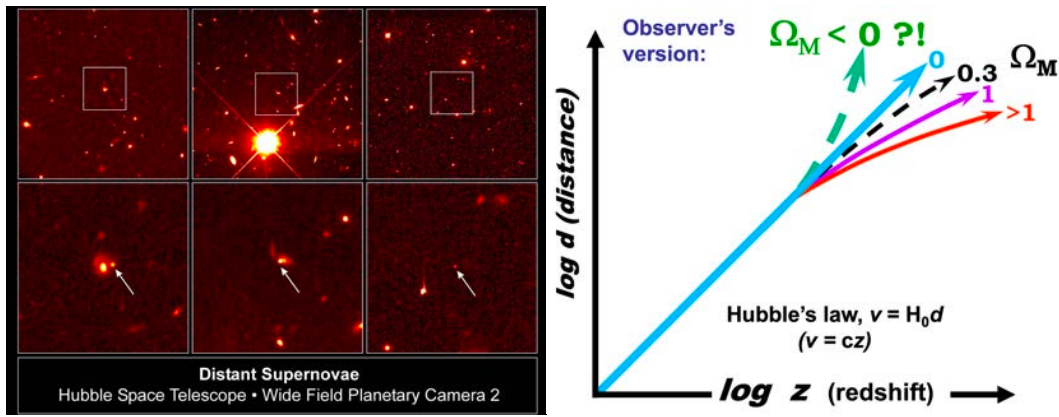

Figure 9 (left): Images of distant galaxies (top) and insets showing three SNe Ia (bottom, marked with arrows). Courtesy NASA/ESA/HST. Figure 10 (right): Same as Figure 4, but now showing where the data actually fell (dashed green curve above the $\Omega_M = 0$ line). But instead of concluding that $\Omega_M < 0$, both teams favored $\Omega_\Lambda > 0$.

So, did our data imply that $\Omega_M < 0$? Just look at Figure 10: $\Omega_M > 1$, 1, 0.3, 0, and < 0. It's a natural conclusion, but we don't know of any anti-gravitating people, or anything like that. And it would have made Einstein kind of unhappy, probably. But in 1917 he had come up with an alternative – which he later supposedly called the "biggest blunder of his life" (or at least of his career) – in order to account for an apparently static universe, in which nearly everyone believed back then. He conjured up this fudge factor $\Lambda$, the "cosmological constant," essentially a constant of integration, but one which he finely tuned to exactly negate the attractive effects of gravity and explain a static universe. A static universe of this sort is mathematically unstable, of course, and it has other problems as well. Why would there be this fine-tuning? Einstein never liked it, but he felt compelled to introduce it in order to explain this apparently static nature of the universe. Indeed, in 1931 he renounced it, when Hubble discovered that the universe isn't static after all.

So instead of saying our data imply $\Omega_M < 0$, let's say that the cosmological constant density of the universe $\Omega_\Lambda > 0$. That would be an alternative explanation. Figure 11 shows the data based on SNe Ia discovered prior to 1998, from Riess et al. (1998) and Perlmutter et al. (1999); this again is proportional to the log of the luminosity distance versus redshift. You can see the points at low redshift, which provide the anchor; those are necessary, otherwise there's nothing with which to compare the high-redshift SNe Ia.

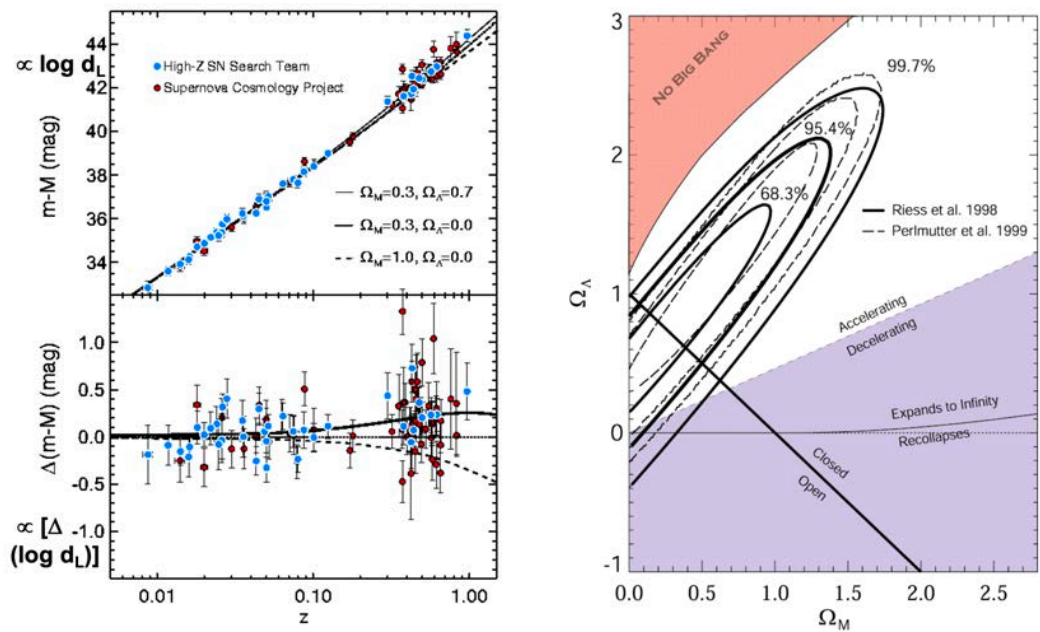

Figure 11 (left): Hubble diagram (top) and residuals relative to the $\Omega_M = 0.3$ and $\Omega_\Lambda = 0$ universe (bottom); data taken from Riess et al. (1998, blue points) and Perlmutter et al. (1999, red points). Figure 12 (right): 1σ, 2σ, and 3σ confidence contours in the $\Omega_M$ vs. $\Omega_\Lambda$ plane, showing that $\Omega_\Lambda > 0$ is most likely according to both teams' data.

At high redshift, the points go above the Einstein-de Sitter universe of $\Omega_M = 1$. And they're even above the low-density ($\Omega_M = 0.3$) preferred universe of the early 1990s. You can see the difference; the data cluster above the $\Omega_M = 0.3$ and $\Omega_\Lambda = 0$ universe. It was only a 2.5–3σ result at the time, far from the 5σ rigors of particle physics – but look, we have one universe and it's hard to gather these data, and if we didn't report this we might have been scooped by someone else. So, astrophysicists can't always rely on 5σ, okay? Sorry about that; we don't have the Large Hadron Collider with which to get trillions of collisions. But the result looked pretty good, you know – high-$z$ data fainter than for a flat or a low-$\Omega_M$ matter universe.

The corresponding $\Omega_M - \Omega_\Lambda$ contours are illustrated in Figure 12. Yes, you can still get an $\Omega_\Lambda$ = 0 universe, but most of the area of these contours is above $\Omega_\Lambda$ = 0. And again, had there been only one team, I don't think people would have taken this very seriously, but two papers reported the results essentially independently: Riess et al. (1998) and Perlmutter et al. (1999). I'll bet Einstein would have been pretty surprised if he were alive in the late 1990s to see his cosmological constant idea resuscitated – not to give a static universe, but one which on large scales is *accelerating*!

Evidence started building up from other sectors as well. The study of large-scale structure – which George Smoot mentioned in his talk – suggested that $\Omega_M$ = 0.3 ± 0.1. The Boomerang cosmic microwave background (CMB) results came out in 2000 (Melchiorri et al. 2000), giving a flat universe; that was enormously useful as well. Looking at the $\Omega_M - \Omega_\Lambda$ diagram (Figure 13), the constraint from large-scale structure (LSS) is a vertical rectangle to a good first approximation (actually a bit tilted), and the contours from the CMB constraint of a roughly flat universe are nearly perpendicular to the SN Ia contours.

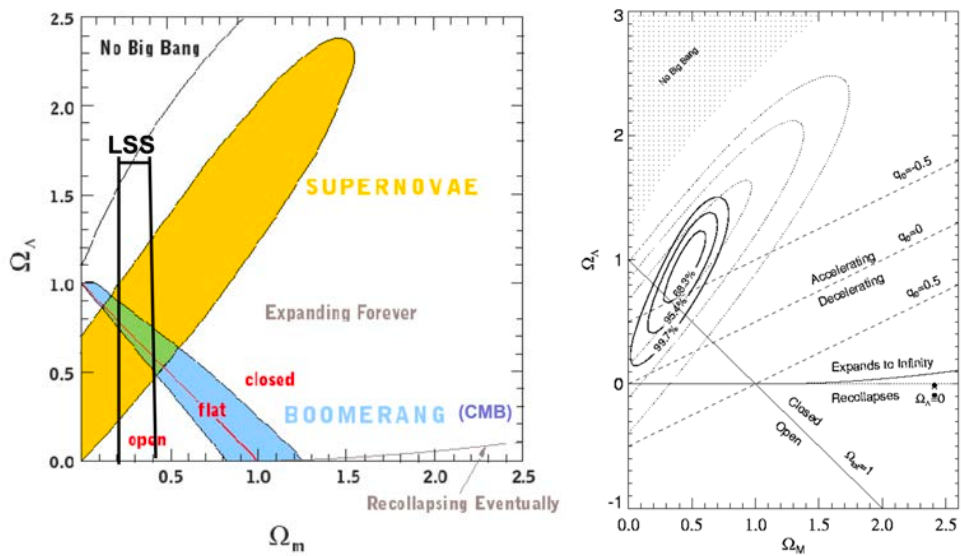

Figure 13 (left): The $\Omega_M$ vs. $\Omega_\Lambda$ plane showing results from SNe Ia, the CMB, and large-scale structure. The three sets intersect around $\Omega_M$ = 0.3, $\Omega_\Lambda$ = 0.7, giving rise to the concordance model. Figure 14 (right): Results from SNe Ia by 2004 (Riess et al. 2004).

So, the "concordance cosmological model" began to emerge and gradually strengthened. This was really pretty radical at the time; hardly anyone liked the cosmological constant. But look, this is what the data seemed to be telling us. The conclusion became even stronger when the beautiful Wilkinson Microwave Anisotropy Probe (WMAP) results appeared – fantastic data!

By around 2004, the SN Ia contours (Figure 14) looked much tighter (Riess et al. 2004). At the 4σ level or so, the supernovae alone are above $\Omega_\Lambda$ = 0, and if you add the constraint that $\Omega_M$ = 0.3, then you're really quite far above the $\Omega_\Lambda$ = 0 line. Also, note that SNe Ia plus large-scale structure gave a precision comparable to that from the CMB plus large-scale structure. In other words, you could say, "I don't like SNe Ia; for some reason they're giving the wrong result." Okay, but if you like large-scale structure and the CMB, you get the *same* result ($\Omega_M$ = 0.3, $\Omega_\Lambda$ = 0.7). We now have three methods, any two of which give the same answer – wonderful! And by this time, an Einstein–de Sitter universe ($\Omega_M$ = 1.0, $\Omega_\Lambda$ = 0) was ruled out at many σ.

We then started probing the expected era of deceleration. If we live in a universe with $\Omega_\Lambda$ = 0.7 and $\Omega_M$ = 0.3, then early in its history it should have been decelerating (Figure 15) because the dark matter would have dominated over any Λ component (or a more general form of repulsive dark energy having constant, or roughly constant, space density), and then it would go through this jerk when it starts accelerating. So, you go way back in time to even higher-redshift SNe Ia than the $z$ = 0.4–0.5 objects upon which our initial results were based. We discovered and studied them with *HST*. Figure 16 is again a plot of log luminosity distance versus redshift, showing the new data (Riess et al. 2007); this has now become a real thing of beauty!

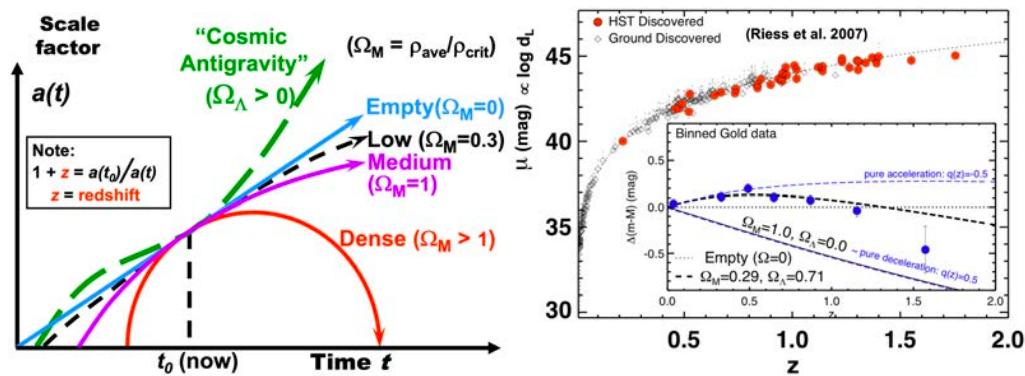

Figure 15 (left): Same as Figure 2, but now also illustrating a model universe in which $\Omega_\Lambda$ > 0. Note the early deceleration of the universe. Figure 16 (right): Data from Riess et al. (2007), showing that SNe I at very high redshifts are brighter than expected if Ω = 0. These results also argue against the unexpected faintness of SNe Ia at $z$ = 0.4–0.5 actually being a consequence of dust or SN evolution.

The inset displays the binned best ("Gold") data relative to this dotted line which represents an empty universe (Ω = 0). You can see the reasonably recent acceleration in the past 5 billion years (SNe Ia are fainter than if Ω = 0), preceded by deceleration (SNe Ia brighter than if $\Omega = 0$). This was fantastic: if the unexpected faintness of the SNe Ia up to redshift 0.5 had been caused by dust or supernova evolution, then you'd expect even *more* dust and even *more* evolution at higher redshifts, so the SNe Ia would look progressively fainter than in an empty universe, whereas if this really is a cosmological feature, you'd expect it to start looking like an Einstein–de Sitter universe ($\Omega_M = 1.0$, $\Omega_\Lambda = 0$) at early times, with the SNe Ia starting to look brighter than if $\Omega = 0$. The data follow the pure acceleration curve in the recent past (a few billion years) and roughly parallel to the pure deceleration curve in the distant past (many billions of years). So, this was a very important test.

Other data supported the presence of $\Lambda$ (or some other form of repulsive dark energy). For example, you can take the wonderful WMAP map that shows temperature variations in the CMB, corresponding to initial cosmological density fluctuations, and run them through a computer simulating the long-term effects of gravity. If you include only visible and dark matter, you end up with large-scale structure that looks kind of like what's in our observed universe, but not quite. But with dark matter *and* repulsive dark energy, the final predicted large-scale structure today ($z = 0$) looks a lot better, close to how our actual observed universe appears. Then there's x-ray clusters of galaxies, the integrated Sachs-Wolfe effect, and other observations; I don't have time to discuss these here, but there's lots of additional supporting evidence.

Figure 17 illustrates the pie chart that emerged; it's a simplified version of the diagram shown by the two previous speakers. A 95% puzzling universe, as Rocky Kolb said. So here we are! Looking at Figure 18, which other speakers have also shown, we've got the birth of the universe for whatever reason, and then quantum fluctuations that are stretched by inflation, leading to the seeds from which large-scale structure later developed. For 9 billion years the expansion decelerated. Then, in the last 5 billion years or so, is has been accelerating – in a sense, a new era of inflation that we have entered. It's nowhere near as rapid, it's not exponential yet, but if the dark energy continues behaving the same way, some day it will become exponential.

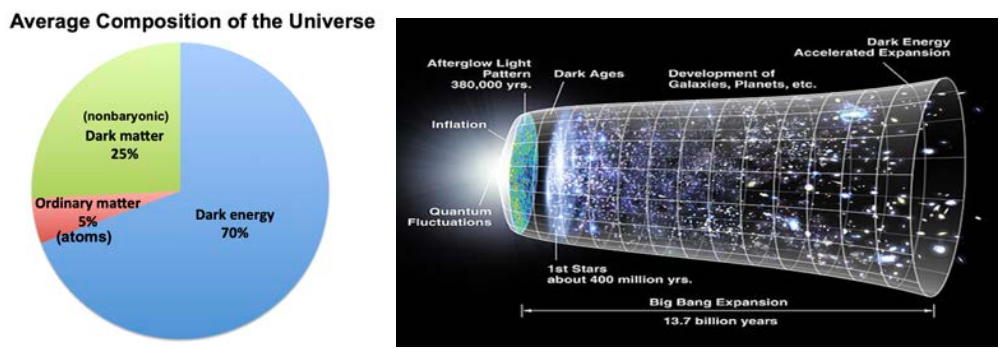

Figure 17 (left): Average composition of the universe. Figure 18 (right): Evolution of the universe. (Credit: NASA)

By 2011, 13 years after the initial announcement of the accelerating universe, there was so much evidence for acceleration that the Nobel Prize in Physics was awarded jointly to Saul Perlmutter, Brian Schmidt, and Adam Riess. They kindly used a significant fraction of their prize money to bring the other team members to participate in Nobel Week; we went to various gatherings and parties. At the official celebratory lunch of the High-$z$ Team, my wife Noelle revealed her "consolation prize" for all of us who did not technically win the Nobel Prize because of the three-people-maximum convention: a limited-edition black t-shirt with a statement written in white, "Dark Energy is the New Black"! She also sent this shirt to the king of Sweden along with a U.S. two-dollar bill, with a note saying, "Here are two items you are unlikely to already have." And in 2015, the Breakthrough Prize in Fundamental Physics was awarded to all team members, recognizing that science these days is often done by teams.

Okay, since dark energy seems real, what is it? As I said, it's not necessarily the cosmological constant, $\Lambda$. Many theorists really don't like the cosmological constant, although I kind of like it. It provides an argument for a multiverse. (I'm sorry, George [Smoot], I favor the multiverse, despite what you said in your lecture!) But many theorists dislike $\Lambda$ because there isn't good agreement between the observations and theory. Theory formally predicts an infinite vacuum energy; sure, if you cut things off at the Planck scale, then it's "only" about $10^{120}$, whereas we measure a very small but nonzero value. And why should $\Omega_\Lambda = 0.7$ right now, close to the value of $\Omega_M = 0.3$, when it changes with time? As Steven Weinberg said, it's a "bone in the throat," this thing that we discovered. So maybe it's not $\Lambda$, maybe it's something else.

We define the equation-of-state parameter, $w = P/\rho c^2$, where $P$ is the pressure and $\rho c^2$ is the energy density. Well, $\rho \propto$ (volume)$^{-(1+w)}$. For normal nonrelativistic matter, $w = 0$; for photons, $w = 1/3$ because they stretch with the expansion of the universe. For $\Lambda$, $w = -1$; in other words, it's a vacuum energy that does not depend on the scale factor, the expansion of the universe. And $w$ is not equal to $-1$ for a whole class of models called quintessence – rolling scalar fields, whatever. As an example, $w = -1/3$ for cosmic strings, which can be ruled out already quite easily. In general relativity, gravitational acceleration is proportional to $-(\rho c^2 + 3P)$. So, all you need is a negative pressure such that $3P$ more than balances $\rho c^2$, and you get a positive acceleration. So that's what you need, $w < -1/3$, and as I said, $w = -1$ for the cosmological constant.

Let's try to measure what $w$ actually is. This is not so easy to do. Figure 19 shows the differences in magnitudes as a function of $z$ for different values of $w$. (For the non-astronomers here, a difference of 0.01 mag corresponds very close to 1% in the brightness of an object, and 0.1 mag is roughly 10%.) So, you have to measure supernova brightness to a few percent in order to distinguish between $w = -1.1$ and $w = -0.9$, for example. That's quite hard to do because there are all kinds of potential systematic effects that can mess up your results. Nevertheless, it's been done, and one relatively recent paper is the joint Sloan Digital Sky Survey (SDSS) and Supernova Legacy Survey (SNLS) analysis (Betoule et al. 2014) shown in Figure 20. Look at this Hubble diagram! Edwin Hubble and Milton Humason would be simply astonished by the quality, by what we've achieved about 85 years after their original Hubble diagrams. They're really beautiful.

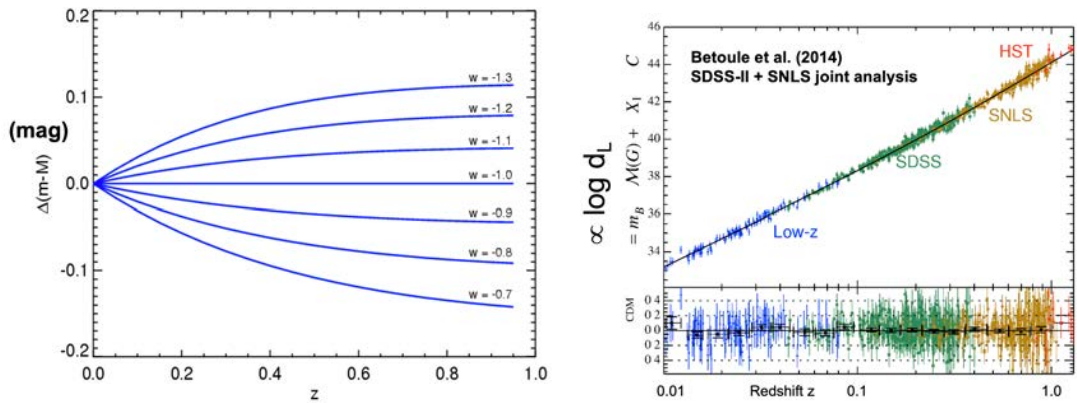

Figure 19 (left): Difference in apparent SN brightness vs. redshift $z$ in an $\Omega_\Lambda$ = 0.7 flat universe. Figure 20 (right): SN Ia Hubble diagram from Betoule et al. (2014); it includes low-$z$ data as an anchor, together with SDSS, SNLS, and *HST* SNe Ia. Residuals relative to the concordance cosmological model are shown at bottom.

From these data, we can plot $\Omega_M$ versus $w$; see Figure 21. Again, we can include constraints from the CMB temperature fluctuations as well as baryon acoustic oscillations. And, lo and behold, there's this pretty good concordance; its contours do intersect $w = -1$, though most of the area is a little bit below that, suggesting a very weird sort of phantom energy whose density grows with time. Now, I don't for a minute believe we've shown the existence of phantom energy; we've got to get much better data for this to be compelling. But in any case, we do have consistency with $w = -1$ and $\Omega_M = 0.3$.

We could assume a time-dependent $w$, and various parameterizations can be used. For example, $w(a) = w_0 + w_a(1 - a)$, where $a = 1/(1 + z)$ is a scale factor that's a function of redshift (Linder 2003); $w_a$ is like a time derivative. For $\Lambda$, $w_0 = -1$ and $w_a = 0$. Okay, so let's do that, and Figure 22 shows what we get (Betoule et al. 2014). Now, we're taking a derivative, so we don't have such good constraints; the diagram has much larger uncertainty contours. But still, it's consistent with $w_0 = -1$ and $w_a = 0$, and thus with $\Lambda$. So far, the null hypothesis that dark energy is the vacuum energy and not a rolling scalar field or something else seems fine. Now, you might say that's a bad null hypothesis because you don't like a vacuum energy, but in a sense it's the simplest hypothesis.

Lately, we've been trying to measure the past expansion history of the universe ever more precisely, in my case with SNe Ia, in order to better predict the future. (There are other complementary measurements which are equally important.) Is dark energy something that's more negative than $w = -1$, in which case we'll experience a "big rip" someday? Kind of a disheartening possibility – we'll all be destroyed, and even the atoms of which you consist will be ripped apart. Again, I don't believe that, but it's possible. Or the dark energy could be constant dark energy; that would be $\Lambda$. Or, perhaps it's some sort of a scalar field whose sign effectively changes in the future. This happened at the end of inflation, so why not again? The inflaton turned into matter that later became us, right? All of us contributed to the Big Bang, something I like to say in my public talks. So, we've made some measurements, and here's our most recent surprise.

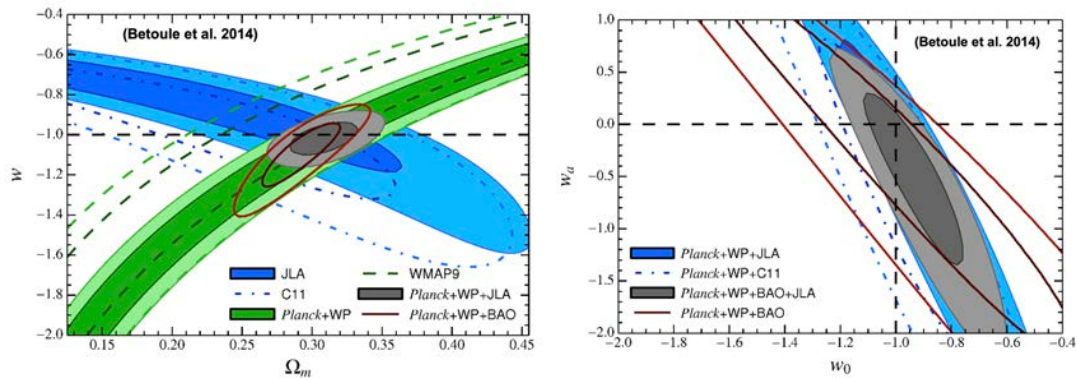

Figure 21 (left): $\Omega_M$ versus $w$, from Betoule et al. (2014). Planck+WP: Planck CMB temperature fluctuations, WMAP CMB polarization. JLA: SNLS-SDSS joint SN Ia light-curve analysis. BAO: baryon acoustic oscillations. The data are consistent with $w = -1$ and $\Omega_M = 0.3$. Figure 22 (right): From Betoule et al. (2014), $w_a$ vs. $w_0$, where $w(a) = w_0 + w_a(1 - a)$ and $a = 1/(1 + z)$, using the joint SDSS and SNLS analysis of SNe Ia. Although the error bars are large, the data are consistent with dark energy being $\Lambda$.

In a set of papers published in the past few years (Riess et al. 2018a,b, 2019, and references therein), the current rate of expansion ($H_0$) might still be too high, even accounting for the Standard Model of cosmology (including $\Omega_\Lambda = 0.7$ and $\Omega_M = 0.3$). What do I mean by this? Okay, you take the beautiful *Planck* CMB map (Figure 23) and construct the angular power spectrum, as in Figure 24. My goodness, look at those data points and the fit; it's incredible. You can define an angular diameter distance as basically the sound horizon length at $t = 380,000$ years divided by the measured angle of the first peak, which corresponds to the angular size of the typical fluctuations. And so, from that, you can predict the current value of the Hubble constant. And here's the result in the Planck collaboration (2016) paper: $H_0 = 66.93 \pm 0.62$ km/s/Mpc (updated to $H_0 = 67.4 \pm 0.5$ km/s/Mpc by Planck collaboration (2018)). Incredibly small error bars!

Previous direct, relatively local measurements by Wendy Friedman and others typically gave numbers in the range $(70–75) \pm (4–7)$ km/s/Mpc. There's a possible conflict with the CMB result, but it's not at all clear; the error bars are big and they themselves are uncertain, and the result is thus not very convincing. So, over a decade ago, Adam Riess and collaborators (including me) launched a campaign to progressively more precisely and accurately measure the current value of the Hubble constant. The final goal is actually to measure $H_0$ to within 1%. We'll achieve that goal, we think, when the final *Gaia* satellite data analysis comes out and with new measurements that we are making. We do it with Cepheids, calibrating their distances in our Milky Way Galaxy more accurately with *HST* and *Gaia*, and using them to determine the distances of the host galaxies of nearby Type Ia supernovae, which are the anchor for the whole SN Ia Hubble diagram.

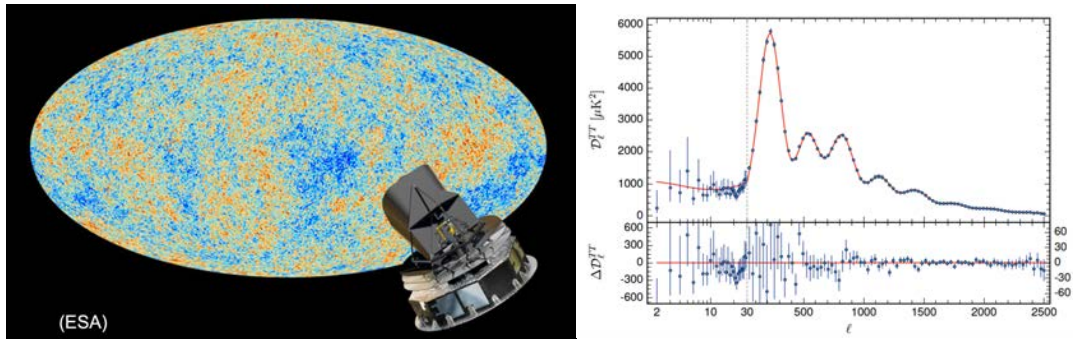

Figure 23 (left): ESA *Planck* satellite map of temperature fluctuations in the CMB. Figure 24 (right): CMB power spectrum from Planck collaboration (2016).

Our latest results (Riess et al. 2018a,b) include some *Gaia* data. *HST* was used to find and measure Cepheid variable stars in the host galaxies of nearby SNe Ia, such as in Figure 25 – they're in the circled regions, pretty faint, but with *HST* you can do amazing things. And so what do we get? Here's our latest result, which strengthens things because we use *Gaia* parallaxes of Milky Way Cepheids. We find that $H_0$ = 73.52 ± 1.62 km/s/Mpc. Compare that with the *Planck* value, 66.93 ± 0.62 km/s/Mpc! Several years ago, we had announced an earlier version of this result, and most astrophysicists said we must be wrong; it doesn't make sense. But we weren't wrong in 1998, when we announced the accelerating expansion of the universe, so maybe we sort of know what we're doing. And even back then we said, "Don't bet your life on it, don't bet your house on it." (In fact, I'm still saying don't bet your *life* on it!)

Nominally there's now a tension of 3.8σ, and we're approaching the particle physics standard of 5σ. [Post-symposium update: Riess et al. (2019) report $H_0$ = 74.03 ± 1.42 km/s/Mpc, bringing the tension to 4.4σ with the Planck collaboration (2018) value of $H_0$ = 67.4 ± 0.5 km/s/Mpc.] We're not there yet, but we're on our way. The uncertainties are smaller than ever before and we think we understand them well. That's the crucial thing: we've examined our uncertainties very carefully, and we think that our uncertainties are reasonable. So maybe this is true!

If you do take $H_0$ = 73.52 km/s/Mpc, $\Omega_\Lambda$ = 0.7, and $\Omega_M$ = 0.3, then in fact the age of the universe is only about 12.8 billion years, not the 13.8 that everyone likes to cite based on a self-consistent *Planck* analysis. (But, by the way, the universe hasn't necessarily been measured by *Planck* at the 1% level. You know, they give these tiny error bars, but… Now, I'm not picking a fight with George Smoot or anyone else; I'm just saying this is a really interesting development that's strengthening with time.)

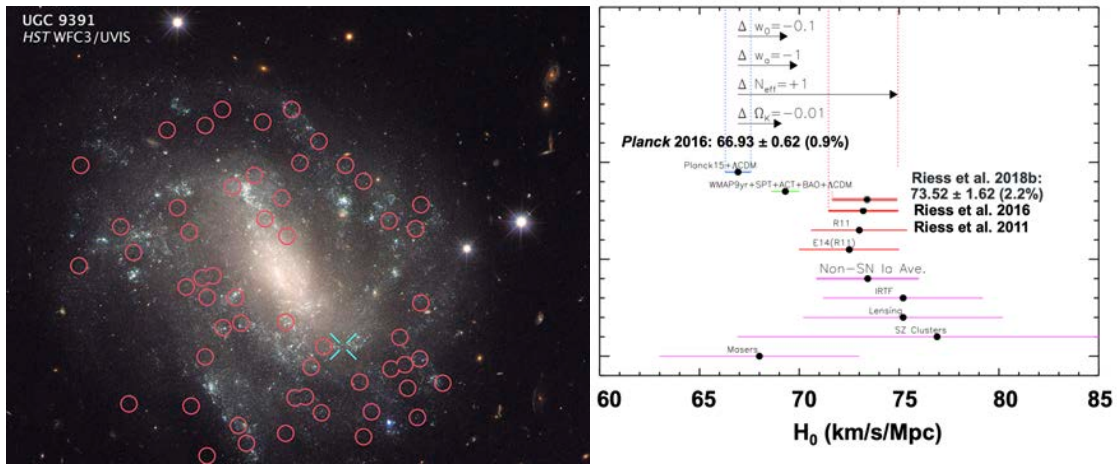

Figure 25 (left): NASA/ESA *HST* image of UGC 9391, the host galaxy of a nearby SN Ia, with detected Cepheids circled. Figure 26 (right): Measurements of $H_0$ from various methods, together with possible ways in which the *Planck* 2016 results could be adjusted to agree with the Cepheid plus SN Ia results of Riess et al. (2019).

Figure 26 shows a set of $H_0$ values from relatively nearby objects: masers, Sunyaev-Zel'dovich effect, gravitational lensing, Tully-Fisher relationship, SNe Ia, and the non-SN Ia average, together with their error bars. And there's the *Planck* result. What could account for this discrepancy (see, e.g., Verdi, Treu, & Riess 2019)? Well, some possibilities include a slightly curved (not flat) universe, or time-variable dark energy, or a change in the effective number of relativistic particles at early times. Perhaps there are more than three neutrino species. Primordial nucleosynthesis and other arguments tell us that there are only three active neutrino species, but maybe there's some sort of a sterile neutrino. That would be pretty cool! General relativity might be wrong, or perhaps there's weird dark matter that interacts with light. Who knows? Or, maybe it's just an error in one or more of our techniques. That would really be boring, but if it's true, we've got to find out someday.

Now, you may dismiss all this, but I'm telling you that over the years the result has become progressively stronger. It was 2.5σ some years ago, and it's now all the way up to 3.8σ [Post-symposium update: 4.4σ], unless we've done something wrong. Well, the way to test it is by using independent methods, to overcome the possible systematics. And the method I want to mention now is very interesting, one with which I have no personal or professional interaction whatsoever: strong gravitational lensing of quasars by galaxies. There are multiple images of a given quasar, and their light paths have different lengths and go through different gravitational potentials. So, this produces nonzero time delays between the brightening and fading of the different images of the same quasar. You can measure the time delays, and you know the geometry from the physics, and thus you get the overall scale and can derive the Hubble constant. This can be done even better with supernovae because they have much more step-like functions in luminosity than quasars do. We plan to utilize this technique in the next few years.

So, here's the result from the H0LiCOW team ($H_0$ Lenses in COSMOGRAIL's Wellspring – a rather tortured acronym, in my opinion!) for three gravitationally lensed quasars (Bonvin et al. 2017, and references therein). They get $H_0$ = 72.8 ± 2.4 km/s/Mpc, consistent with our SN Ia result but with somewhat larger error bars, though they're getting progressively smaller. [Post-symposium update: Birrer et al. (2019) report $H_0$ = 72.5 (+ 2.1, –2.3) km/s/Mpc from four strongly lensed quasars, and Wong et al. (2019) obtain $H_0$ = 73.3 ± 1.8 km/s/Mpc from a set of six quasars; combined with the SN Ia results, the tension with *Planck* reaches $5\sigma$. The review by Riess (2020) reveals even greater overall tension, even $6\sigma$, when the results of additional techniques are included.] And again, with strongly-lensed supernovae this technique will become even better. So, they basically agree with us: a local value of the Hubble constant that differs from the early-time CMB data and their extrapolation.

My own currently favorite explanation for the early/late discrepancy is that it's some sort of a new particle that the CMB researchers haven't yet taken into account when using the Standard Model to make their prediction, because previously there was no reason to do so. But now possibly there's a reason to do so. I find this to be a more palatable explanation than dark energy becoming stronger with time, or dark matter that somehow interacts with photons, and certainly I find it more attractive than general relativity being wrong. Anyway, we don't know the answer yet. But start paying more attention to this problem – it's only growing with time. Stay tuned!

I'd like to thank the conference organizers for inviting me to speak, Frank Serduke for drawing some of the diagrams, and the various agencies and individuals that have funded my research (NSF, NASA, DoE, AutoScope Corporation, TABASGO Foundation, Sylvia and Jim Katzman Foundation, Gary and Cynthia Bengier, Christopher R. Redlich Fund, Richard and Rhoda Goldman Fund, UC Berkeley Miller Institute for Basic Research in Science). Thank you very much!

**[Note added prior to posting on arXiv, December 2023:** The "Hubble tension" between the CMB-based prediction and the current measurements with Type Ia supernovae, discussed at the end of the lecture, has now grown to $5\sigma$: Riess et al. (2022, ApJL, 934, L7; arXiv:2112.04510) report $H_0$ = 73.04 ± 1.04 km/s/Mpc, and an extension of that work (Murakami et al. 2023, JCAP, 046, 33; arXiv:2306.00070) refines the result to $H_0$ = 73.29 ± 0.90 km/s/Mpc. In a very important study, Riess et al. (2023, ApJL, 956, L18; arXiv:2307:15806) recently showed with *James Webb Space Telescope (JWST)* near-infrared observations of two nearby galaxies that the previous *HST* Cepheid photometry was *not* affected by crowding; the resulting dispersion in the Cepheid period-luminosity relations is decreased by a factor of >2.5, thereby improving the precision of individual Cepheids from 20% to 7%. There is no systematic difference in distances obtained with *HST* and *JWST* observations of Cepheids. Moreover, other techniques for measuring the current value of the Hubble constant have also been improving and give results consistent with those from SNe Ia. For example, Blakeslee et al. (2021, ApJ, 911, 65; arXiv:2101.02221) measured the distances of 63 bright galaxies within 100 Mpc with infrared *HST* observations of surface brightness fluctuations, finding $H_0$ = 73.3 ± 0.7 (stat.) ± 2.3 (sys.) km/s/Mpc. To me, it seems unlikely that the late-time determination of $H_0$ (from several methods) is erroneous. Though not an expert on the CMB-based technique, I am inclined to believe the work of my CMB colleagues, so I conclude that the Hubble tension is real. Something is probably missing from the standard cosmological model, perhaps an early (pre-CMB) episode of dark energy – but all solutions proposed thus far have difficulties. The field is ripe for progress, with many new observations being conducted or planned. As I stated at the end of my June 2018 lecture, "Stay tuned!"

**[Covault]** Thank you, Alex, that was great. We have some time for a question or two.

**[Filippenko]** There's a bunch of them, so you choose.

**[Covault]** We'll start, oh okay, we'll start there.

**[Q&A Question 1]** Hi. So, I know this question might be kind of impossible to answer, but everyone always is, like, saying, oh, like, the universe is expanding and stuff, but nobody really says what it's expanding into.

**[Filippenko]** Yeah.

**[Question 1]** I was just wondering if you had any, like, theories or anything to talk about.

**[Filippenko]** Yeah, what is the universe expanding into? It might not be expanding into anything. You can have a mathematically expanding universe; its coordinate system expands according to general relativity. But if that's too abstract for you, think of it as expanding into another dimension in a bigger hyperspace. The example I like to give is a balloon, a two-dimensional closed universe. If you allow the laws of physics to only operate within the rubber of the balloon, it's a two-dimensional universe – you can go forward and backward, left and right, or any combination of those two motions – but you're not allowed to go in or out, yet it's expanding in that radial outward direction, into a bigger hyperspace, this room. So, you can think of it that way, but mathematically you don't have to assert the actual physical existence of that other dimension. It doesn't have to expand into anything, just like the real-number line can expand but it doesn't expand into anything. It remains the same infinity.

**[Question 2]** I was just wondering, there's still multiple scenarios for what exactly causes a Type Ia supernova. Does that ever keep you up at night?

**[Filippenko]** Yes, the issue that we don't truly understand the progenitor, and to some degree the explosion mechanism, of a Type Ia supernova sometimes does keep me up at night. We're making progress on that front. But at the very least we can say that, if we use these empirical standardizations and all that, the Hubble diagram dispersion goes down at small redshifts and at high redshifts. Whenever dispersions go down, you build some confidence, but it'll be better once we truly understand the progenitors. I actually think there are several mechanisms, both the so-called single-degenerate model and the double-degenerate model, and maybe even the dynamical interaction in a three-body system where two white dwarfs actually collide. But we're working on that and it's a very important project.

**[Q&A Question 3]** What about the Hubble constant measurement from gravitational wave events?

**[Filippenko]** Yeah, yeah. So, as I said here, we want as many methods as possible to independently test the results. This, historically, is why the accelerating universe eventually became accepted by most cosmologists – not because of only Type Ia supernovae, but because of the concordance model.

So, the "standard sirens" of neutron-star merging pairs – fantastic, right? From the wave form you know exactly what kind of system merged, and unlike the inverse-square law of light where it's diluting over the surface of a sphere, the amplitudes of gravitational waves are inverse-distance. Since you know what the initial amplitude was, and you measured an amplitude with LIGO/Virgo, that gives you a distance based on this siren. And you measure the redshift of the host galaxy in the usual way from its spectrum. Thus, you've got exactly what you need: a distance measurement that's independent of Hubble's law, so that you can measure the Hubble constant. So that'll be a way of doing it. And the very first event, the one that we saw in August 2017, gives a Hubble constant that's pretty much consistent with anything reasonable – not with what de Vaucouleurs would have said 50 years ago, that it's 100 km/s/Mpc – but it's something like 70 plus or minus 10 or 12. So, that one system is not yet very constraining, but when LIGO/Virgo turns on again in less than a year, they should start finding more of these neutron-star merging pairs, and that'll be a new method with which to measure distances.

**[Covault]** Great, thanks. Thank you, that was a great talk. Let's thank Alex one last time.